\documentclass[%
,twocolumn% 
,secnumarabic%
,amssymb, amsmath,nobibnotes, aps, prb]{revtex4}

\usepackage{bm}%
\usepackage[dvips]{graphicx, color}

\usepackage{comment}
 
\expandafter\ifx\csname package@font\endcsname\relax\else
 \expandafter\expandafter
 \expandafter\usepackage
 \expandafter\expandafter
 \expandafter{\csname package@font\endcsname}%
\fi

\begin{document}
\title{Effect of co-doping of donor and acceptor impurities in the ferromagnetic semiconductor 
Zn$_{1-\textit{x}}$Cr$_\textit{x}$Te studied by soft x-ray magnetic circular dichroism }

\author{Y.~Yamazaki}
\email{yamazaki@wyvern.phys.s.u-tokyo.ac.jp}
\affiliation{Department of Physics, University of Tokyo, 
Bunkyo-ku, Tokyo 113-0033, Japan}

\author{T.~Kataoka}
\affiliation{Department of Physics, University of Tokyo, 
Bunkyo-ku, Tokyo 113-0033, Japan}

\author{V.~R.~Singh}
\affiliation{Department of Physics, University of Tokyo, 
Bunkyo-ku, Tokyo 113-0033, Japan}

\author{A.~Fujimori}
\affiliation{Department of Physics, University of Tokyo, 
Bunkyo-ku, Tokyo 113-0033, Japan}

\author{F.-H.~Chang}
\affiliation{National Synchrotron Radiation Research Center (NSRRC), Hsinchu 30076, Taiwan, Republic of China}

\author{D.-J.~Huang}
\affiliation{National Synchrotron Radiation Research Center (NSRRC), Hsinchu 30076, Taiwan, Republic of China}

\author{H.-J.~Lin}
\affiliation{National Synchrotron Radiation Research Center (NSRRC), Hsinchu 30076, Taiwan, Republic of China}

\author{C.~T.~Chen}
\affiliation{National Synchrotron Radiation Research Center (NSRRC), Hsinchu 30076, Taiwan, Republic of China}

\author{K. Ishikawa}
\affiliation{Institute of Materials Science, University of Tsukuba, 1-1-1 Tennoudai, Tsukuba, Ibaraki 305-8573, Japan}

\author{K. Zhang}
\affiliation{Institute of Materials Science, University of Tsukuba, 1-1-1 Tennoudai, Tsukuba, Ibaraki 305-8573, Japan}

\author{S. Kuroda}
\affiliation{Institute of Materials Science, University of Tsukuba, 1-1-1 Tennoudai, Tsukuba, Ibaraki 305-8573, Japan}

%\date{\today}

\begin{abstract}
We have performed x-ray absorption spectroscopy (XAS) and x-ray magnetic circular dichroism (XMCD) studies of the diluted ferromagnetic semiconductor
Zn$_{1-\textit{x}}$Cr$_\textit{x}$Te doped with iodine (I) or nitrogen (N), corresponding to electron or hole doping, respectively.
From the shape of the Cr $2p$ absorption peak in the XAS spectra, it was concluded that Cr ions in the undoped, I-doped and lightly N-doped samples are divalent (Cr$^{2+}$), while Cr$^{2+}$ and trivalent (Cr$^{3+}$) coexist in the heavily N-doped sample. This result indicates that the doped nitrogen atoms act as acceptors but that doped holes are located on the Cr ions. In the magnetic-field dependence of the XMCD signal at the Cr $2p$ absorption edge, ferromagnetic behaviors were observed in the undoped, I-doped, and lightly N-doped samples, while ferromagnetism was considerably suppressed in heavily N-doped sample, which is consistent with the results of magnetization measurements.
\end{abstract}

%\pacs{75.50.Pp,78.30.Fs,75.30.Hx,78.70.Dm}
%\keywords{diluted magnetic semiconductor, x-ray magnetic circular dichroism, photoemission spectroscopy, electronic structure}

\maketitle
\section{Introduction}
Diluted magnetic semiconductors (DMSs) \cite{furdyna(1988)} which show high ferromagnetic Curie temperatures ($\textit{T}_\textrm{c}$'s) are considered to be key materials for future applications of spintronics~\cite{ohno(1998),ohno(2002)}. The II-IV DMS Zn$_{1-\textit{x}}$Cr$_\textit{x}$Te is known to show ferromagnetism at room temperature, as confirmed by magnetization measurements and magnetic circular dichroism (MCD) measurements in the visible to ultra-violet region \cite{saito(2003)}. 
Recently, the effects of additional doping of atoms with different valencies in Zn$_{1-\textit{x}}$Cr$_\textit{x}$Te were investigated, that is, iodine which is expected to act as an n-type dopant enhances the ferromagnetism~\cite{ozaki(2006)}, while nitrogen which is expected to act as as a p-type dopant suppresses it~\cite{ozaki(2005)}. 
On the other hand, it has been predicted theoretically that spinodal decomposition causes the apparent ferromagnetic behavior of Zn$_{1-\textit{x}}$Cr$_\textit{x}$Te~\cite{sato(2005),fukushima(2006)}.
Experimentally indeed, spatially resolved energy-dispersive x-ray spectroscopy revealed that the Cr ions are distributed inhomogeneously in the I-doped samples while they are distributed homogeneously in the N-doped samples~\cite{kuroda(2007)}. 
The results of $\textit{ab initio}$ calculations of the total energies suggest that the valence state of Cr seems to be important for distribution of the Cr ions,
that is, Cr ions will be distributed inhomogeheously if Cr ions are in the neutral 2+ charge state, while they are distributed homogeneously if Cr ions are in different +$(2\pm \delta)$ charge states~\cite{dietl(2006), kuroda(2007)}.
 So far, x-ray magnetic circular dichroism studies have been done on Zn$_{1-\textit{x}}$Cr$_\textit{x}$Te to investigate  the electronic structure of Cr ions and Cr ions have been found to be in the 2+ state~\cite{ishida(2008), kobayashi(2008)}. It is necessary to know how the electronic state of the Cr ion is modified by the N- and I- doping to understand the mechanism of the spinodal decomposition.
  In order to investigate the effects of I- and N-doping,
we have performed x-ray absorption spectroscopy (XAS) and x-ray magnetic circular dichroism (XMCD) experiments at the Cr 2$\textit{p}$ absorption edge of the undoped, I-doped, lightly N-doped, and heavily N-doped Zn$_{1-\textit{x}}$Cr$_\textit{x}$Te thin films.
XMCD is defined as the difference in XAS spectra between right-handed ($\mu^{+}$) and left-handed ($\mu^{-}$) circularly polarized x-rays, and is a powerful tool to investigate about ferromagnetism in DMS \cite{hwang(2005),mamiya(2006),song(2008),takeda(2008),kobayashi(2008a)} because it is element specific and 
sensitive to magnetically active components.

\section{Experimental}
The samples used in this study were undoped Zn$_{1-\textit{x}}$Cr$_\textit{x}$Te with $x=0.053$ ($\textit{T}_\textrm{c}=90$ K), I-doped Zn$_{1-\textit{x}}$Cr$_\textit{x}$Te with $x=0.039$ ($\textit{T}_\textrm{c}=210$ K), lightly N-doped Zn$_{1-\textit{x}}$Cr$_\textit{x}$Te with $x=0.043$ ($\textit{T}_\textrm{c}=60$ K), and heavily N-doped Zn$_{1-\textit{x}}$Cr$_\textit{x}$Te with $x=0.047$ (no ferromagnetism). 
According to secondary ion mass spectroscopy (SIMS) analysis, the N concentration in the lightly and heavily N-doped samples was
$1.8 \times 10^{18} \textrm{cm}^{-3}$ and $1.0 \times 10^{20} \textrm{cm}^{-3}$, respectively.
These values correspond to the ratios of substituting N for Te of 0.1 \% and 0.56 \%, 
respectively, when we assume all the N atoms substitute for Te.
 These samples were grown on insulating GaAs (001) substrates by molecular beam epitaxy, as described elsewhere~\cite{kuroda(2007)}. After depositing a 600 nm thickness of ZnTe buffer layer, the Zn$_{1-x}$Cr$_x$Te thin films of 300 nm thickness were grown. During the deposition, the substrate was kept at a temperature of 603-633 K. 
The sample surface was capped with a 3 nm thick Al layer to avoid oxidazation for the Zn$_{1-\textit{x}}$Cr$_\textit{x}$Te layer.
XAS and XMCD measurements were performed at the Dragon Beamline BL11-A of National Synchrotron Radiation Research Center (NSRRC), Taiwan. 
Spectra were measured both in the total-electron-yield (TEY) mode and the total-fluorescence-yield (TFY) mode. Because the probing depth of the TEY and TFY mode are about 5 nm and 100 nm, respectively, one can consider that the TEY mode is surface sensitive and the TFY mode is bulk sensitive. 
The monochromator resolution was $E/\Delta E > 10,000$. In the XMCD measurements, the circular polarization of the incident photons was fixed and the direction of the applied magnetic field was changed. The XAS and XMCD measurements were made at a temperature of 20 K in an ultrahigh vacuum below $\sim$ $10^{-10}$ Torr.
  
\section{Results and Discussion}

\begin{figure}
\includegraphics[width=8cm]{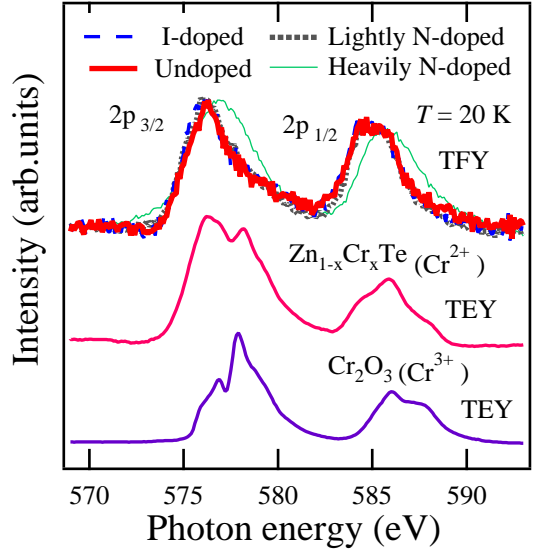}
\caption{(Color online) Comparison of the XAS spectra of the undoped, I-doped, lightly N-doped, and heavily N-doped Zn$_{1-\textit{x}}$Cr$_\textit{x}$Te thin films taken in the TFY mode. The XAS spectra of undoped Zn$_{1-\textit{x}}$Cr$_\textit{x}$Te and Cr$_2$O$_3$ taken in the TEY mode are shown as references of Cr$^{2+}$ and Cr$^{3+}$, respectively.}
\label{XAS}
\end{figure}

Figure \ref{XAS} shows the Cr 2$\textit{p}$ XAS spectra of the undoped, I-doped, lightly N-doped, and heavily N-doped Zn$_{1-\textit{x}}$Cr$_\textit{x}$Te thin films taken in the TFY mode. For comparison, those of Cr$_{2}$O$_{3}$ and Zn$_{1-\textit{x}}$Cr$_\textit{x}$Te taken in the TEY mode are also shown. 
Cr ions in Cr$_{2}$O$_{3}$ and Zn$_{1-\textit{x}}$Cr$_\textit{x}$Te are Cr$^{3+}$ and Cr$^{2+}$, respectively, as described elsewhere~\cite{theil(1999), kobayashi(2008)}.
The major two peaks in each spectrum are due to the $2p_{3/2}-2p_{1/2}$ spin-orbit doublet of the Cr 2$\textit{p}$ core level. 
Due to self-absorption effect of the TFY method, the intensity of 2$\textit{p}_{3/2}$ peak relative to the 2$\textit{p}_{1/2}$ peak is reduced compared to one obtained by the TEY method.
One can see in Fig.\ref{XAS} that the spectral line shapes and the peak positions of the undoped, I-doped and lightly N-doped samples are similar to the spectra of Cr$^{2+}$. Considering that the TFY mode is bulk sensitive, one can conclude that Cr$^{2+}$ is dominant in the undoped, I-doped and lightly N-doped samples in the bulk region. 
On the other hand, the XAS spectra of the heavily N-doped sample showed clear peak shifts towards higher energies compared to the ohter samples and located between  Cr$^{2+}$ and Cr$^{3+}$. This observation indicates that N replacing Te atoms in the heavily N-doped sample supplied holes and works as an acceptor  and that Cr$^{3+}$ states are created.

\begin{figure}
\includegraphics[width=8.5cm,clip]{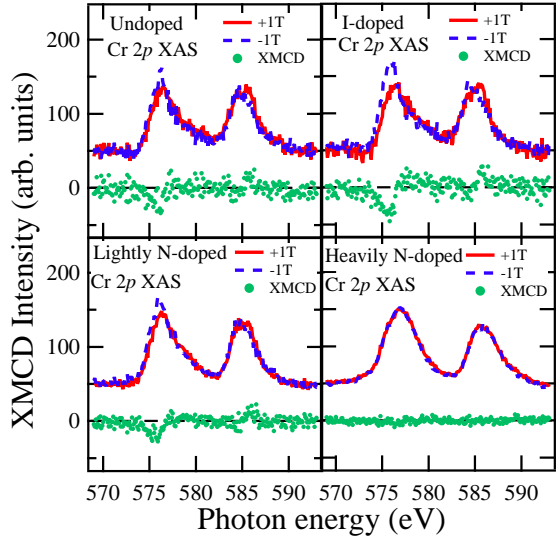}
\caption{(Color online) Cr 2$\textit{p}$ XAS and XMCD spectra of the undoped, I-doped, lightly N-doped, and heavily N-doped Zn$_{1-\textit{x}}$Cr$_\textit{x}$Te thin films taken by the total fluorescence yield mode at $\textit{H}$ = 1T and $\textit{T}$ = 20 K. The XMCD spectra have been normalized to the
 XAS $[(\mu^{+}+\mu^{-})/2]$ peak height at around 576 eV.}
\label{XASandXMCD}
\end{figure}

\begin{figure}
\includegraphics[width=8.5cm,clip]{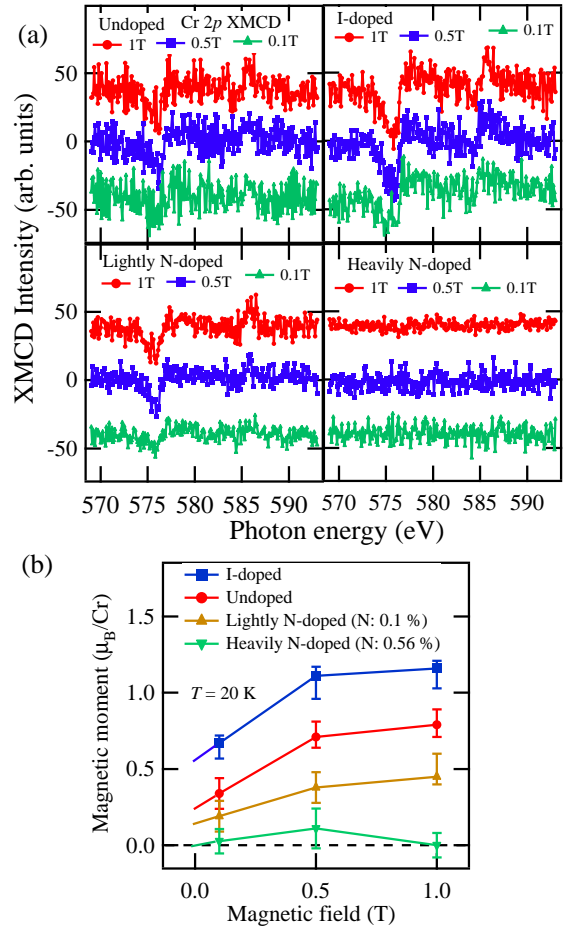}
\caption{(Color online) Magnetic field dependence of XMCD spectra (a) and magnetic moments (b) of the I-doped, undoped, lightly N-doped, and heavily N-doped Zn$_{1-\textit{x}}$Cr$_\textit{x}$Te thin films at $\textit{T}$ = 20 K.}
\label{XMCDandMM}
\end{figure}

Figure \ref{XASandXMCD} shows the Cr 2$\textit{p}$ XAS and XMCD spectra of the undoped, I-doped, lightly N-doped, and heavily N-doped Zn$_{1-\textit{x}}$Cr$_\textit{x}$Te thin films taken at $\textit{T}$ = 20 K. Magnetic fields were applied perpendicular to the sample surfaces. We observed clear XMCD signals in the undoped, I-doped and lightly N-doped samples, while the heavily N-doped sample showed no clear XMCD signals, consistent with the magnetization measurements~\cite{kuroda(2007)}.
Figure \ref{XMCDandMM}(a) shows the XMCD spectra of the undoped, I-doped, lightly N-doped, and heavily N-doped samples taken under 1, 0.5 and 0.1 T.
 The line shapes of the XMCD spectra are same for all the samples, indicating Cr$^{2+}$ ions contribute to the magnetism in these samples. 
Although applying of XMCD sum rules~\cite{carra(1993), thole(1992)} to TFY data does not give accurate results because of the self-absorption effect,
we have attempted to deduce the magnetic moment of Cr ions by applying the XMCD sum rules to the TFY data in order to see relative changes with N or I doping.
Figure \ref{XMCDandMM}(b) shows the magnetic field dependence of the approximate Cr magnetic moments of these samples estimated using the XMCD sum rules. 
If the XMCD intensities are extrapolated to $H$ = 0 T, one can see finite magnetization at $\textit{H}$ = 0 T in the undoped, I-doped, and lightly N-doped samples, indicating that the ferromagnetism in these samples are of intrinsic bulk origin. In addition, a higher magnetic moment was observed in the I-doped sample than in the undoped sample. 
Although the application of the XMCD sum rules to TFY data does not give accurate magnetic moments,
the observation of finite magnetization at $\textit{H}$ = 0 T and the relative changes of magnetization with N or I doping are real one and can be used to study the ferromagnetism of the undoped, I-doped, and lightly N-doped samples and the influence of N or I doping on the magnetic state of Cr.

The recent atomic-scale analysis using energy-dispersive x-ray spectroscopy has shown a close correlation between the spatial homogeneity of Cr distribution with ferromagnetic properties~\cite{kuroda(2007)}; the inhomogeneous distribution, possibly due to spinodal decomposition, enhances the ferromagnetism while the homogeneous distribution suppresses it. According to a theoretical model~\cite{dietl(2006)}, the valence state of Cr ions is important for the spinodal decomposition. In the intrinsic situation, the valence state of Cr ions substituting for Zn is 2+, being electrically neutral with the same charge state as Zn$^{2+}$, and strong attractive interaction between the electrically neutral Cr ions makes the Cr distribution inhomogeneous. When the Cr charge state deviates from 2+, the repulsive force of the electrostatic origin created between the Cr ions makes a homogeneous distribution. In the actual samples grown by MBE, it is considered that the Cr valence state in the undoped Zn$_{1-\textit{x}}$Cr$_\textit{x}$Te slightly increases  from 2+ due to the formation of Zn vacancies, which act as shallow acceptors, while the Cr valence returns to 2+ in the I-doped Zn$_{1-\textit{x}}$Cr$_\textit{x}$Te due to the compensation of these Zn vacancies by the donor impurities. This model explains the experimentally observed inhomogeneous/homogeneous Cr distribution and the resultant enhancement/suppression of ferromagnetism in the I-doped/undoped Zn$_{1-\textit{x}}$Cr$_\textit{x}$Te. In either case, the dominant Cr ions are in the valence state of 2+, which is consistent with the result of the XAS measurements in the I-doped and undoped samples. On the other hand, the co-doping of nitrogen, which acts as an acceptor impurity when substituting for Te, converts the Cr valence state from 2+ to 3+. As a result, in the heavily N-doped sample, the Cr valence state becomes the mixed valence or an intermediate valence between Cr$^{3+}$ and Cr$^{2+}$, as demonstrated in the XAS spectra. The disappearance of ferromagnetism in the heavily N-doped sample can be attributed to this conversion of the Cr valence state in addition to the homogeneous Cr distribution. In the lightly N-doped sample, the conversion of the Cr valence is not sufficient to induce an apparent shift of the XAS spectra.
Our experiment thus gives the first microscopic support of the model that acceptor doping changes the valence of Cr and this change influences the ferromagnetism in Zn$_{1-\textit{x}}$Cr$_\textit{x}$Te.

\section{Conclusion}
In conclusion, we have performed XAS and XMCD measurements on undoped, I-doped, lightly N-doped, and heavily N-doped Zn$_{1-\textit{x}}$Cr$_\textit{x}$Te. From the XAS measurements, Cr ions were found to be in the Cr$^{2+}$ state in the undoped, I-doped and lightly N-doped samples, while Cr$^{3+}$ and Cr$^{2+}$ states were formed in the heavily N-doped sample. In the XMCD measurements, it was found that the magnetic moment of the Cr$^{2+}$ state contributes to the ferromagnetism in the undoped, I-doped and lightly N-doped samples, while the XMCD signals at the Cr $2p$ edge were not observed in the heavily N-doped sample. The magnitude of magnetic moment was found to be larger in the I-doped sample than in the undoped one, consistent with the results of the magnetization measurements.

\section{Acknowlegement}
The SIMS measurement was performed by K. Hayama at the AIST Nano-Processing Facility, supported by "Nanotechnology Network Japan" of the Ministry of Education, Culture, Sports, Science and Technology (MEXT), Japan.
This work was supported by a Grant-in-Aid for Scientific Reserch in Priority Area gCreation and Control of Spin Currenth (19048012) form the Ministry of Education, Culture, Sports, Science and Technology, Japan. TK is supported in part by a Global COE Program "the Physical Sciences Frontier", MEXT, Japan.

\clearpage

\end{document}